# Enabling Social Internet of Things and Social Cloud

Weishan Zhang 1, Qun Jin 2, Didier El Baz 3
1 Department of Software Engineering, China University of Petroleum
No.66 Changjiang West Road, Qingdao, China. 266580
2 Department of Human Informatics and Cognitive Sciences, Waseda University, Japan
3 LAAS-CNRS, 7 avenue du Colonel Roche 31077, Toulouse Cedex 4, France

**Abstract**—Social Internet of Things are changing what social patterns can be, and will bring unprecedented online and offline social experiences. Social cloud is an improvement over social network in order to cooperatively provide computing facilities through social interactions. Both of these two field needs more research efforts to have a generic or unified supporting architecture, in order to integrate with various involved technologies. These two paradigms are both related to Social Networks, Cloud Computing, and Internet of Things. Therefore, we have reasons to believe that they have many potentials to support each other, and we predict that the two will be merged in one way or another.

## 1 INTRODUCTION

With the development of Internet of Things (IoT), people begin to live in a cyber-physical-social hyper space. The Internet of Things are changing dramatically what it means to be social. The very obvious examples are smart phones, tablets, laptops and all other wearable devices are connecting
people directly or indirectly through different applications, like Twitter1, Facebook2, WeChat3, and other social network applications. Everyday objects and people are connected with each other to establish social relationships and form a social network, which is called Social Internet of Things (SIoT) [1]. SIoT will bring unprecedented online and offline social experiences. SIoT is the interconnection of different services, objects and people where participants can benefit from involving in this network. For example, smart transportations that involving vehicle networks, drivers, smart phones, and other devices may make traffic congestion as less as possible where traffic information are timely disseminated at real time. As the next evolutionary step of the internet of things, SIoT is expected to enhance performance of object discovery, service composition, and evaluation of trustworthiness of objects [2]. However, even there are some projects that are dedicated to provide SIoT platforms (e.g the project Social- IoT 4 and INPUT5), it is noted that SIoT enabling technologies need more efforts in different directions including [2]:
- Definition of inter-object relationships, like semanticrepresentational models for social relationships,  solutions to autonomously sense other objects and exchange information.
- Network analysis algorithms tailored for SIoT after the establishment of a social network.
- Architecture tailored for SIoT that allows for establishment and exploitation of social relationships, resources and services.

A closely related field with respect to SIoT is the "Social Cloud (SoC)" [3] [4], which is an improvement over social network in order to cooperatively provide computing facilitiesthrough social interactions. SoC is playing important roles in different areas, e.g. serving as shared infrastructure
for software crowdsourcing [5]. Another interesting example is a SoC supported approach for helping to alleviate communication breakdowns led by asymmetries in media and time preferences among different ages of family members [6]. SoC and SIoT can be integrated together to make use of advantages of both paradigms. Novel data fusion algorithms, artificial intelligence techniques can be used to realize automated decision making in SoC in order to support better communication and collaboration among people and things in SIoT. The combination of SoC and SIoT can realize ubiquitous sensing and computing beyond the capability of individual of people or things, and stimulate innovations in various fields.

## 2 WHAT DO WE COVER IN THIS SPECIAL ISSUE?

This special issue presents most recent research on SoC and SIoT. We have three papers on various research issues of SIoT in which one paper presents their work for integrating SoC with SIoT. There is one paper that presents a review of SoC research. SIoT can be used to enable the integration of a variety of devices into people daily life, being empowered by the interconnectivity and the FOAF (Friend of a Friend) feature offered by Social Networking Services (SNS). As a promising paradigm, SIoT can provide an infrastructure to study and integrate the intelligence mechanisms required to enhance services adaptability and user-friendliness. In their article "Social Cloud-based Cognitive Reasoning for Task-oriented Recommendation in the Social Internet of Things", Dina Hussein, Son N. Han, Gyu Myoung Lee and Noel Crespi from Institut Mines-Telecom and Liverpool John Moores University describe their approach to the integration of cognitive reasoning into SIoT for providing recommendation ofquotidian tasks in smart homes, in which reasoning about both physical and social aspects of context is required to achieve situation characterization. They propose and developan intelligent recommendation framework as a service built on top of SoC, by utilizing the reasoning mechanism on context elements represented by ontologies. ThingsChat, a proof-of-concept prototype, is built and explained with experiment results, which demonstrates improvement in adaptability of recommendation results to users situations. Implementations of the SIoT model envision cyber counterparts of physical objects are called social virtual objects (SVOs), virtualized in the cloud. However, most of the IoT devices do not have the processing and communication capabilities required to create and manage social relationships. The paper "Social Virtual Objects in the Edge Cloud: a key component for distributed IoT applications" by I. Farris, R. Girau, L. Militano, M. Nitti, L. Atzori, A. Iera and G. Morabito, investigates how to address this issue by exploiting computing resources in the edge of a network to hostSVOs of the SIoT. Edge cloud technologies is used for the implementation of their SIoT platform, which is expected to reduce latency, to increase scalability, and to ease management of physical node mobility. Through experimental simulations, they investigate the frequency of each type of message exchanged under different inter-cloud migrations of processes associated to virtual objects, and assess the latency reduction w.r.t. conventional cloud solutions. There have been three types of efforts to enable SoC since it was proposed in 2010 [3], namely Social Compute Cloud supports sharing of compute resources between friends, Social Storage Cloud supports storage resource sharing, allowing users to store data on friends resources, Social Content Delivery Network (S-CDN) provides a fabric for replicating and sharing data, using friends resources as intermediate content delivery nodes, as summarized in the paper titled "Social Clouds: A Retrospective" by Kyle Chard, Simon Caton, Omer Rana, Kris Bubendorfer. Based on their previous implementations, the authors propose a general social cloud architecture to facilitate heterogeneous bilateral resource sharing. Interoperability and cloud federation standards thus will be important to ensure seamless resource usage. Similar to SIoT research, suitable algorithms and metrics are needed to analyze social clouds. In Social Internet of Things (SIoT), devices are usually divided to groups with different relationship in social networks. However, radio access network (RAN) used for connecting IoT devices is short of supporting SIoT groups. For this, network virtualization with software defined network (SDN) structure is a scalable solution. But this solution is hard to support enough groups due to rule space limitation of existing SDN enabled devices. In their work titled "Cloud Radio Access Network Virtualization for Social Internet of Things" by He Li, Mianxiong Dong, and Kaoru Ota, a SDN based RAN virtualization framework is proposed to maximize the number of SIoT groups with limited SDN rule space. Extensive simulations show that the proposed solution provides more SIoT groups compared with some other allocation methods, with better performance at thesame time.

## 3 CONCLUSION

Both SIoT and SoC are at infant stage, and more research efforts are need to have a better, and probably unified supporting architecture, in order to integrate with heterogeneous technologies involved in SIoT and SoC. Security and trust issues are still there and obviously need more attentions, especially in social environments. And clearly, SIoT and SoC have more potentials to go ahead hand by hand, and to support each other. We would like to say that the two have very good potentials to merge as a unified paradigm.

## ACKNOWLEDGMENTS

This special issue would not have been possible without the support of Mazin Yousif, editor in chief of IEEE Cloud Computing, colleagues from the IEEE Computer Society office, and all reviewers involved the review process.

**Weishan Zhang** Weishan Zhang is a full professor, deputy head for research of Department of Software Engineering, School of Computer and Communication Engineering, China University of Petroleum (UPC). He is the director of 'Big Data Intelligent Processing Innovation Team of Huangdao District', director of 'Big Data Processing for Petroleum Engineering of UPC', and founding director of the Autonomous Service Systems Lab. He was a Research Associate Professor/Senior Researcher at Computer Science Department, University of Aarhus (til Dec. 2010). He was a visiting scholar of Department of Systems and Computer Engineering, Carleton University, Canada (Jan. 2006 - Jan. 2007). He was an Associate Professor at School of Software Engineering, Tongji University, Shanghai, China (Aug. 2003- June 2007). He was a NSTB post-doctoral research fellow at Department of Computer Science, National University of Singapore (Sept. 2001 to Aug. 2003). His current h-index according to google schoolar is 14 and the numberof total citations is around 540 in Dec. 2015.

**Qun Jin** un Jin is a professor and the chair of the Department of Human Informatics and Cognitive Sciences, Waseda University, Japan. His recent research interests cover human-centric ubiquitous computing, behavior and cognitive informatics, big data, personal analytics and individual modeling, e-learning, e-health, and computing for wellbeing.Jin received a PhD in computer science and electrical engineering from Nihon University. He is a member of IEEE. Contact him at jin@waseda.jp.

**Didier El Baz** Dr. Didier El Baz is a CNRS research scientist at LAAS-CNRS since 1985. He is head and founder of the Distributed Computing and Asynchronism team (CDA). His H index is 21 and I10 index is 30. He has broad research interests including Parallel computing, distributed computing, Peer to peer computing, Applied Mathematics and Optimal Control.